# Trees Assembling Mann-Whitney Approach for Detecting Genome-wide Joint Association among Low-Marginal-Effect loci

Changshuai Wei[1], Daniel J. Schaid[2], Qing Lu[1*]

[1]Department of Epidemiology and Biostatistics, Michigan State University, East Lansing, MI, 48824 USA

[2]Division of Biostatistics, Harwick 7, Mayo Clinic, 200 First Street Southwest, Rochester, Minnesota 55905, USA.

*Correspondence to: Qing Lu, Department of Epidemiology and Biostatistics, Michigan State University; B601 West Fee Hall, East Lansing, Michigan, 48824, USA. Email: qlu@epi.msu.edu

# Abstract

Common complex diseases are likely influenced by the interplay of hundreds, or even thousands, of genetic variants. Converging evidence shows that genetic variants with low-marginal-effects (LME) play an important role in disease development. Despite their potential significance, discovering LME genetic variants and assessing their joint association on high-dimensional data (e.g., genome-wide association studies) remain a great challenge. To facilitate joint association analysis among a large ensemble of LME genetic variants, we proposed a computationally efficient and powerful approach, which we call Trees Assembling Mann-Whitney (TAMW). Through simulation studies and an empirical data application, we found that TAMW outperformed multifactor dimensionality reduction (MDR) and the likelihood-ratio-based Mann-Whitney approach (LRMW) when the underlying complex disease involves multiple LME loci and their interactions. For instance, in a simulation with 20 interacting LME loci, TAMW attained a higher power (power=0.931) than both MDR (power=0.599) and LRMW (power=0.704). In an empirical study of 29 known Crohn's disease (CD) loci, TAMW also identified a stronger joint association with CD than those detected by MDR and LRMW. Finally, we applied TAMW to Wellcome Trust CD GWAS to conduct a genome-wide analysis. The analysis of 459K single nucleotide polymorphisms was completed in 40 hours using parallel computing, and revealed a joint association predisposing to CD (p-value=2.763e-19). Further analysis of the newly discovered association suggested that 13 genes, such as *ATG16L1* and *LACC1,* may play an important role in CD pathophysiological and etiological processes.

# Introduction

With the rapid advances in high throughput technologies, genome-wide association studies (GWAS) have been widely undertaken to study the genetic causes of common complex diseases. The increasing number of associated loci discovered from these studies will likely shed light on the underlying disease pathophysiological and etiological processes, as well as promote genome-based translational research. Nevertheless, a substantial proportion of the genetic variants that contribute to common complex diseases remain uncovered[Eichler, et al. 2010].

Studies suggest that genetic variants with low-marginal-effects (LME) could contribute to a significant proportion of disease heritability. For instance, a GWAS on human height found LME single nucleotide polymorphisms (SNPs) could account for nearly half of the total genetic variance[Yang, et al. 2010]. Despite their potential importance, LME SNPs are generally associated with moderate significance levels, and thus are rarely detected by a single locus approach using a stringent p-value threshold (e.g., 5e-8). Complex diseases are also likely influenced by numerous LME SNPs. For example, hundreds of genetic variants are estimated to be involved in Crohn's Disease[Park, et al. 2010]. Identifying such a large number of genetic variants and considering their possible interactions poses a great challenge to genetic association research.

Statistical approaches have been developed to search for joint associations among multiple genetic variants, with or without considering interactions. In this paper, we focused on joint association approaches allowing for possible interactions. Among these approaches,

Multifactor Dimensionality Reduction (MDR) has been widely used[Ritchie, et al. 2001]. MDR exhaustively searches all of the possible subsets of SNPs to find the best combination that divides individuals into a "high-risk" group and a "low-risk" group. In this way, MDR reduces dimensionality, making it suitable for a joint association analysis. However, the computation time required for an exhaustive search increases exponentially with the number of SNPs, making it infeasible to apply the method directly to genome-wide data.[Cordell 2009] We had previously developed a likelihood-ratio-based Mann-Whitney (LRMW) approach for genome-wide joint association analysis allowing for interactions. The approach adopts a fast Mann-Whitney (MW) based forward selection algorithm[Ye, et al. 2011] to search the entire genome for disease susceptibility SNPs, and assesses their joint association using an LRMW test. However, the LRMW is conservative in selecting disease susceptibility SNPs, and tends to only detect joint associations among a limited number of strong-marginal-effect (SME) SNPs (e.g., less than 10 SNPs). This makes it less ideal for evaluating complex joint association involving hundreds or thousands of LME SNPs.

To facilitate a joint association analysis on a large number of LME SNPs while allowing for possible interactions, we propose a Trees-Assembling Mann-Whitney approach (TAMW). TAMW uses a trees-assembling technique[Breiman 2001; De Lobel, et al. 2010; Jiang, et al. 2009; Lunetta, et al. 2004; Schwarz, et al. 2010] to search the genome for disease-susceptibility SNPs, including LME SNPs, and then evaluates their joint association using an MW test. Through simulation studies and a real data application, we systematically evaluated the performance of TAMW, and compared it with MDR and LRMW. The new approach was then applied to Wellcome Trust Crohn's Disease GWAS CD dataset of 459,091

SNPs for a genome-wide joint association analysis.

# Methods

In TAMW, we build multiple tree models from bootstrap samples of original data and then combine these models into a trees-assembling model. We then evaluate the joint association, as well as individual contributions, of genetic variants selected into the assembling model (including a large number of LME genetic variants) using an MW test.

## Trees-Assembling Mann-Whitney Approach

To illustrate the approach, we assume there are $N$ individuals in the data. Each individual is genotyped with $P$ SNPs. We first draw $T$ bootstrap samples from the original data. For each bootstrap sample, we randomly select $S$ SNPs from $P$ SNPs, and then build a tree model by applying a forward selection algorithm[Ye, et al. 2011] to the selected SNPs. The forward selection algorithm starts with a null model without any SNPs, and then gradually selects SNPs to form multi-locus genotype groups. In step one, the algorithm searches $S$ SNPs for a disease-susceptibility SNP to divide samples into two genotype groups (e.g., a group of individuals carrying *AA* genotype and the other group of individuals carrying *Aa/aa* genotypes), which gives the highest possible MW statistic. In step two, we search for the second SNP, considering its possible interaction with the first SNP, to split the two existing groups into four two-locus groups, associated with the highest possible Mann-Whitney statistic. A special case may occur when the same SNP is selected to further split the two genotype groups. In this particular case, the model contains only three genotype groups (i.e., *AA*, *Aa*, and *aa*) at step two. The whole splitting process continues until the number of multi-locus genotypes reaches a pre-defined size. The idea of forward selection algorithm is inherited from the classic tree approach. However, it differs from tree in two aspects: 1) in each step, it only allows a single SNP to split existing genotype groups so that joint association model identified from forward selection algorithm is more consistent with existing models (e.g., multiple-interaction model[Marchini, et al. 2005]); 2) it adopts MW statistic as the model selection criteria, which facilitate the statistical significance assessment of selected joint association model, as demonstrated below.

Assuming a total of $M_t$ multi-locus genotypes in a tree model $t$ ($1 \leq t \leq T$), we can calculate the likelihood ratio ($LR_m^t$) for each multi-locus genotype ($G_m^t$):

$$LR_i^t = \frac{P(G_m^t \mid D)}{P(G_m^t \mid \bar{D})} \quad 1 \leq m \leq M_t,$$

where $D$ and $\bar{D}$ denote disease status and non-disease status, respectively. By repeatedly applying the forward selection algorithm to $T$ bootstrap samples, we obtain an ensemble of tree models. Each tree model assigns an LR value ($LR_i^t, 1 \leq i \leq n$) to an individual $i$ ($1 \leq i \leq n$) according to one's multi-locus genotype (e.g., $LR_i^t = LR_m^t$, if $i$-th individual carries $m$-th multi-locus genotype). The assembling LR value ($LR_i^{TA}$) for an individual $i$ can thus be obtained by averaging the LR values over $T$ tree models ($LR_i^{TA} = \frac{1}{T} \sum_{t=1}^{T} LR_i^t$).

Based on the ranks of the LR values, we can form the Trees-Assembling Mann-Whitney statistic to evaluate the joint association of the selected SNPs with the disease

$$U = \sum_{i=1}^{N_D} \sum_{j=1}^{N_{\bar{D}}} \psi(LR_i^{TA}, LR_j^{TA}), \tag{1}$$

where $N_D$ and $N_{\bar{D}}$ represent the number of cases and non-cases. The kernel function is defined as

$$\psi(LR_i^{TA}, LR_j^{TA}) = \begin{cases} 1, & \text{if } LR_i^{TA} > LR_j^{TA} \\ 0.5, & \text{if } LR_i^{TA} = LR_j^{TA} \\ 0, & \text{if } LR_i^{TA} < LR_j^{TA} \end{cases}.$$

By assembling a large number of tree models, each comprised of different sets of SNPs, the new approach could simultaneously consider a large number of LME SNPs and their interactions. Randomly selecting a subset of SNPs that are not likely to be comprised of any SME SNPs allows TAMW to also increase the chances of considering important interactions among LME SNPs.

**Evaluating the joint association**

Given the above Mann-Whitney statistic, hypothesis testing can be formed to test the joint association:

$$Z = \frac{(U - U_0)}{\sqrt{Var(U)}}, \qquad (2)$$

where $U_0$ equals $0.5 N_D N_{\bar{D}}$. $Var(U)$ can be calculated as

$$Var(U) = S_D + S_{\bar{D}}, \qquad (3)$$

where

$$S_D = \sum_{i=1}^{N_D} \left( \sum_{j=1}^{N_{\bar{D}}} \psi(LR_i^{TA}, LR_j^{TA}) - \frac{U}{N_D} \right)^2,$$

$$S_{\bar{D}} = \sum_{j=1}^{N_{\bar{D}}} \left( \sum_{i=1}^{N_D} \psi(LR_i^{TA}, LR_j^{TA}) - \frac{U}{N_{\bar{D}}} \right)^2.$$

Under the null hypothesis, Z follows a standard normal distribution asymptotically, which can be used to estimate the significance level of the detected joint association. However, model selection, ranking LR, and performing hypothesis testing on the same data could inflate Type I error. Although permutation test could be used to adjust the inflated significance level, it is not computationally feasible for a genome-wide joint association analysis, especially when the analysis involves complicated modeling (e.g., the assembling procedure). Alternatively, we can evaluate the joint association of the selected disease-susceptibility SNPs on an independent dataset (e.g., a replication dataset or a split sample). To test the joint associations on an independent dataset, we first apply the $T$ tree models to the independent dataset, so that each subject is assigned $T$ LR values. We then average the $T$ LR values to form a TAMW statistic. Based on the TAMW statistic, we evaluate the significance of the joint association using equation (2).

## Individual contributions of genetic variants

Importance measurements (IM) can be used to assess the individual contributions of genetic variants in the joint association. The IM is measured by comparing the joint association model with an interested SNP to the model without the SNP.

Assuming that a specific SNP $s$ has been selected in the $T_s$ tree models, we can calculate the IM for each tree model $t$ ($1 \leq t \leq T_s$) using a Z score,

$$Z_t^s = \frac{\Delta U_t^s}{\sqrt{Var(\Delta U_t^s)}} = \frac{U_{t1}^s - U_{t0}^s}{\sqrt{Var(U_{t1}^s) + Var(U_{t0}^s) - 2Cov(U_{t1}^s, U_{t0}^s)}}, \qquad (4)$$

where $U_{t1}^s$ and $U_{t0}^s$ are the MW values for the tree model $t$, with and without the SNP $s$,

respectively. $Var(U_{t1}^s)$ and $Var(U_{t0}^s)$ are the corresponding variances, which can be calculated based on equation (3). $Cov(U_{t1}^s, U_{t0}^s)$ can be calculated as:

$$Cov(U_{t1}^s, U_{t0}^s) = V_D^{s,t} + V_{\bar{D}}^{s,t}, \qquad (5)$$

where,

$$V_D^{s,t} = \sum_{i=1}^{N_D}\left\{(\sum_{j=1}^{N_{\bar{D}}}\psi(LR_i^{s,t1}, LR_j^{s,t1}) - \frac{U_{t1}^s}{N_D})(\sum_{j=1}^{N_{\bar{D}}}\psi(LR_i^{s,t0}, LR_j^{s,t0}) - \frac{U_{t0}^s}{N_D})\right\},$$

$$V_{\bar{D}}^{s,t} = \sum_{j=1}^{N_{\bar{D}}}\left\{(\sum_{i=1}^{N_D}\psi(LR_i^{s,t1}, LR_j^{s,t1}) - \frac{U_{t1}^s}{N_{\bar{D}}})(\sum_{i=1}^{N_D}\psi(LR_i^{s,t0}, LR_j^{s,t0}) - \frac{U_{t0}^s}{N_{\bar{D}}})\right\}.$$

Given the IM($Z_t^s$) for each tree model $t$, the individual contribution of the SNP $s$ to the joint association can be measured by averaging IM values from all of the $T_s$ tree models ($Z_s = \frac{1}{T_s}\sum_{t=1}^{T_s} Z_t^s$).

## Results

### Simulation Studies

Through simulation studies, we evaluated the TAMW approach, and compared its performance with two existing approaches, MDR and LRMW. We first conducted a simple simulation to study the performance of three approaches under complex disease scenarios involving multiple LME loci, and their two-way and three-way interactions. To mimic a real disease scenario, we simulated the genotype data from the Wellcome Trust Crohn's disease (CD) data, and introduced a combination of independent LME loci, two-way interactions and three-way interactions into the disease model. In the second simulation, we evaluated the performance of the three approaches under five different types of gene-gene interaction models, varying from those with marginal effects to those with limited or no marginal effect. In both simulations, we started with simple disease models including only one gene-gene interaction, and then gradually increased the model's complexity by including additional independent LME loci or interactions.

For each simulation, we evaluated both type I error and power of the approaches. Power (type I error) is defined as the probability of identifying a significant joint association (i.e., p-value <0.05) when there is an (no) association. For TAMW and LRMW, we used 2/3 of the data to search for a joint association model, and then assessed the significance of the joint association model on the remaining 1/3 data using the MW based test. For MDR, we adopted the same procedure used in previous MDR papers[Moore, et al. 2007; Pattin, et al. 2009]by implementing a Tuned ReliefF filter to exclude noise SNPs, and then applying MDR on the remaining SNPs to search for a joint association. The significance of the joint association can be then obtained based on an extreme value distribution.[Pattin, et al. 2009]

**Simulation I**

The genotype data of simulation was generated based on 29 CD-associated SNPs from the Wellcome Trust CD study. To evaluate the type I error, we first simulated a null model comprised of no functional SNP (i.e., all 29 SNPs are noise SNPs). The power was then assessed under five disease models, comprised of 2, 5, 10, 15 and 20 SNPs, respectively. The first model represented a simple disease scenario with only one two-way interaction. As the number of functional SNPs increased, we gradually increased model complexity by introducing additional single-locus, two-way interaction and three-way interaction effects. At the same time, we decreased the effect-sizes of the functional SNPs so that each of them was associated with a lower marginal effect. The details of the simulation settings were summarized in Table S1. The rationale behind the simulation was to evaluate the performance of the three approaches under a variety of disease models. While the first few models mimicked a simple disease caused by few SME SNPs, the later models were closer to a complex disease that involved a large number of LME SNPs and their interactions.

We applied TAMW, MDR and LRMW on the simulated data, and calculated the type I error and power based on 1000 simulated replicates (Table I). The results showed that the type 1 error from all three approaches was well controlled at the level of 0.05. For the simplest model with one two-way interaction, TAMW (Power=0.691) was comparable with MDR (Power=0.705) and LRMW (Power=0.812). As the number of functional SNPs increased and their effect-size decreased, TAMW demonstrated its advantage over the other two. When the number of disease-susceptibility SNPs was greater than 10, TAMW attained the highest power among the three approaches. For the most complicated model of 20 LME SNPs, the power of TAMW reached 0.931, much higher than MDR (Power=0.599) and LRMW (Power=0.704).

**Simulation II**

Statistical approaches may perform differently under various underlying gene-gene interaction models. For instance, MDR has a unique feature of capturing interactions with no marginal effect, while LRMW attains higher power when at least one of the interacted markers has a marginal effect. In this simulation, we considered five two-way interaction models. The first three interaction models were a multiplicative effect model (M1), a multiplicative interaction model (M2) and a threshold interaction model (M3), respectively[Marchini, et al. 2005]. These models were commonly used in previous literatures and assumed marginal effects on both loci. The fourth interaction model (M4) came from a typical MDR simulation, assuming no marginal effect on either loci[Ritchie, et al. 2003]. The fifth interaction model (M5) represented another possible disease model with marginal effect on only one locus (Table S4). Similar to simulation I, we included 29 SNPs in the simulation data. First, a null model with no functional loci was simulated to assess the type I error. We then evaluated the power of the three approaches under each of five two-way interaction models. For each interaction model, we started with a simple disease model with one interaction between two LME loci, and gradually added LME loci to increase the model's complexity, until a complex model was reached with five two-way interactions. For simplicity, we only included the same type of interaction into a model and assumed they were associated with same effect-size.

We applied TAMW, MDR and LRMW to the simulated data. The type I error for TAMW, LRMW and MDR were 0.055, 0.042 and 0.042, respectively, which were well controlled at the level of 0.05. The three approaches had comparable power under the simple disease model with one interaction (Figure 1). While TAMW and LRMW were slightly more powerful than MDR under the interaction models with marginal effects (i.e., M1, M2, M3, and M5), MDR obtained more power than TAMW and LRMW when the interacted loci had no marginal effect (i.e., M4). However, as the number of interacting SNPs increased, the performance of TAMW improved significantly under interaction models with marginal effects, while the performance of LRMW and MDR only obtain limited improvement.

TAMW attained the highest power under the complex model with five interactions. For example, under the disease model with five threshold interactions (i.e., M3), the power of TAMW was much higher (Power=0.854) than the power of MDR (Power=0.276) and the power of LRMW (Power=0.378). Under the interaction model with no marginal effects (i.e., M4), all three approaches obtained a limited power increase as the number of LME SNPs increased.

**Application to Crohn's Disease**

Crohn's disease (CD) is a major form of inflammatory bowel disease[Hanauer 2006]. Genetics contribute significantly to CD, with an estimated sibling relative risk between 25 and 42.[Russell and Satsangi 2004] However, known CD genetic variants only explain a small proportion of CD heritability. Franke et al[Franke, et al. 2010] recently conducted a meta-analysis and identified 71 CD-associated SNPs, explaining 23.2% of the total genetic variance. CD is likely caused by the interplay of hundreds or thousands genetic variants[Park, et al. 2010]. Although recent GWAS and meta-analysis have revealed numerous genetic loci strongly related to CD, a large proportion of LME loci remain to be discovered. The identification of these LME loci and their possible interactions should not only help elucidate how LME genetic variants interplay with each other within biological pathways to cause disease, but also explain additional heritability.

For this purpose, we conducted a joint association analysis by using a large-scale Wellcome Trust GWAS data[Burton, et al. 2007], which is comprised of 1748 unrelated cases and 2938 unrelated controls. A candidate gene analysis was carried out to evaluate joint associations among 29 known CD-associated SNPs[Barrett, et al. 2008]. Using the candidate gene data, we also compared the performance of TAMW, MDR and LRMW. Moreover, in order to uncover unknown LME genetic variants and their interactions, we carried out a genome-wide joint association analysis by applying TAMW to a total of 459,091 SNPs.

**Joint Association Analysis among known loci**

We applied TAMW, MDR and LRMW to the Wellcome Trust CD dataset to evaluate joint associations among 29 known CD-associated SNPs. TAMW completed the analysis in 70 seconds on a personal computer with two 2.50GHz CPUs and 3.25 GB of memory, which was comparable to LRMW (31 seconds) but much faster than MDR (1500 seconds). With the capacity to consider both SME SNPs and LME SNPs (Table S3), TAMW identified a joint association highly associated with CD (p-value=1.84e-39). On the other hand, MDR and LRMW only detected joint associations among 3 SME SNPs, which attained significance levels of 5.81e-05 and 6.25e-14, respectively (Table II). To gain an idea how TAMW would perform on SME SNPs alone, we applied TAMW to the 3 SNPs selected by MDR and LRMW. The joint association detected among these 3 SME SNPs by TAMW reached a comparable significance level (p-value=3.33e-16) with the one identified by LRMW.

Importance measurement analysis followed the initial evaluation, to assess the individual contribution of the 29 SNPs to the joint association detected by TAMW. The analysis revealed the significant contributions of *IL23R*, *ATG16L1*, *LACC1*, *TNFSF15* and *NKX2-3* to the identified joint association. Among these five genes, *IL23R* and *ATG16L1* were ranked as the most important genes ($1^{st}$ and $2^{nd}$), which is consistent with the results of single locus analysis (Table S3), as well as the outcomes from joint association analyses using MDR and LRMW (Table II). Aside from the evidence from our study, previous studies have repeatedly reported *IL23R* and *ATG16L1* as being associated with CD[Cummings, et al. 2007; Duerr, et al. 2006; Okazaki, et al. 2008]. The remaining 3 top-ranked genes (i.e., *LACC1*, *TNFSF15* and *NKX2-3*) were not detected by MDR and LRMW. Yet, TAMW suggested that they could make a substantial contribution to the genetic susceptibility of CD, which is consistent with published literatures[Franke, et al. 2010; Umeno, et al. 2011; Yamazaki, et al. 2005; Yu, et al. 2009]. Yamazaki et al[Yamazaki, et al. 2005] reported that *TNFSF15* was associated with CD in both Japanese and European populations. Further, a cytology experiment indicated that the gene might function via the innate immune system to protect the intestinal barrier. Yu et al[Yu, et al. 2009] found that *NKX2-3* gene expression is up-regulated in CD patients and suggested that the abnormal expression of the gene might alter gut migration of antigen-response and thus increase susceptibility to CD. The *LACC1* gene was found to be associated with CD in multiple studies[Franke, et al. 2010; Umeno, et al. 2011], but its function in the etiology of CD remains unknown.

## Genome-wide Joint Association Analysis

We conducted a genome-wide joint association analysis to search for novel gene variants, especially those LME variants, associated with CD. For the genome-wide joint association analysis, we simultaneously evaluated 459,091 SNPs that had passed the Wellcome Trust quality control criteria. We applied TAMW to two-thirds of the Wellcome Trust CD samples to search for a parsimonious joint association model associated with CD, and then applied the model to the remaining one-third of samples to assess its significance level. Due to the intensive computation, we performed the genome-wide joint association analysis on a clustered machine. We paralleled the computations onto 20 CPUs (2.3GHz), with each CPU assigning 4 GB of memory. The joint association model was converged at about 2000 trees (Figure S1) and a total of 20,000 trees were grown to ensure that all of the SNPs were covered[Lunetta, et al. 2004; Schwarz, et al. 2010].

The genome-wide analysis was completed in 40 hours, which identified a joint association significantly associated with CD (p-value=2.763e-19). To gain better insight into the joint association, we assessed the individual contribution of each SNP through their IM values. We focused on the 100 highest IM-ranked SNPs (Table S4), from which we identified thirteen genes (Table III). Among the thirteen genes, seven of them (i.e., *NOD2*, *IL23R*, *ATG16L1*, *PTPN2*, *CYLD*, *C5orf56*, and *SLC22A5*) had previously been reported as associated with CD [Barrett, et al. 2008; Ferguson, et al. 2007; Franke, et al. 2010]. In our analysis, these seven genes were consistently top-ranked in both single locus analysis (via PLINK, [Purcell, et al. 2007]) and the joint association analysis (via TAMW). For the remaining six genes, *LACC1, ZGPAT*, *TNFSF15*, *IL12RB2*, *GALNT2*, and *MCF2L2*, single locus analysis showed that they were only moderately associated with CD, while joint association analysis indicated they might interact to play an important role in CD disease etiology.

In fact, previous studies have confirmed the association of three of these six LME genes (*LACC1, ZGPAT* and *TNFSF15*) with CD. Franke et al[Franke, et al. 2010] detected the association of *LACC1* and *ZGPAT* with CD in a meta-analysis study based on 6333 cases and 15056 controls. In another GWAS study of a Japanese population, Yamazaki et al[Yamazaki, et al. 2005] revealed that *TNFSF15* was associated with CD. Although no previous study had reported the association of *IL12RB2* and *GALNT2* with CD, they have plausible biological links to CD. *IL12RB2* interacts with *IL23R* in the *IL23* signaling pathway—a critical regulator in CD[Wang, et al. 2009]. *GALNT2* might function in maintenance of the mucin layer by interacting with *NOD2*, which could influence susceptibility to CD[Phillips, et al. 2009]. *MCF2L2* has been reported as a susceptibility loci in Type 1 and Type 2 diabetes[Takeuchi, et al. 2008; Zhang, et al. 2010]; our study suggests it may also relate to CD.

## Discussion

Despite the recent success of genome-wide association studies (GWAS), a large proportion of the genetic variations associated with complex diseases remain unknown. Research has revealed the important role of LME genetic variants in disease pathophysiology and etiology. Nevertheless, the majority of the LME genetic variants are understudied because of their moderate association with diseases. Hundreds or perhaps even thousands of LME genetic variants lie on the genome. Although individual LME genetic variant has limited effect on diseases, they could act jointly to account for a significant part of the unexplained variance. We developed a new approach, TAMW, to evaluate the joint association of a large number of LME genetic variants and their potential interactions with a disease. Through simulation studies, we were able to show that TAMW had advantages over two existing approaches, LRMW and MDR, when the disease involved multiple LME genetic variants and their interactions. We also conducted an empirical study by applying TAMW to Wellcome Trust CD GWAS data with nearly 500K SNPs. The genome-wide analysis was completed in 40 hours. The identified SNPs were then ranked based on their IM values, from which we found thirteen genes predisposing to CD. Previous genetic association studies have suggested that these genes play important roles in the etiology of Crohn's disease. Yet, among the thirteen genes, six of them were associated with low marginal effect, and would likely be missed in a single locus analysis. Compared with a single-locus analysis that only considers marginal effects, the joint association analysis takes both marginal effect and interaction effect into account, and thus provides a more comprehensive assessment of genetic variants' role in disease association.

Compared to existing approaches, TAMW has four unique features: 1) TAMW could simultaneously consider a large number of LME SNPs and their interactions through assembling multiple tree models. Although each tree model is moderately associated with disease outcomes, combining them together could result in great improvement in association. 2) By randomly selecting a subset of genetic variants, likely comprised of only LME SNPs, TAMW also increases the chance of identifying important interactions among the LME SNPs. 3) TAMW is well designed for computational efficiency with the capacity of handling GWAS data. It could easily parallel the computation and build the forward trees simultaneously on a cluster machine. 4) The results from TAMW can be interpreted in several ways. From TAMW, we can assess the significance of a joint association with a disease. Moreover, a strong joint association also indicates its potential for risk prediction. In fact, MW statistic can be easily transformed to AUC ( $AUC = \dfrac{U}{N_D N_{\bar{D}}}$ ) to measure the prediction ability of multiple genetic markers. The one-to-one relationship between MW statistic and AUC also links the MW statistic to disease heritability. As shown by So et al[So and Sham 2010], heritability could be inferred from AUC by knowing the disease prevalence.

The search algorithm in TAMW is developed based on the concept of ensemble learning, which has been adopted in other ensemble learning approaches, such as Random Forest (RF). However, unlike the commonly used classification and prediction approach, RF, TAMW is an approach designed for genetic association analyses, and has several unique features. Firstly, TAMW is built under the framework of MW statistic, which facilitates the association test of multiple disease-susceptibility markers with the disease. Secondly, instead of CART, TAMW uses a forward selection algorithm to build model. As we demonstrated elsewhere[Lu, et al. 2012], the forward selection algorithm is computationally more efficient, making TAMW feasible for genome-wide association analyses. TAMW forms a Mann-Whitney statistic on the rank of LR values rather than actual LR values. Similar as other non-parametric approaches, it could be slightly less powerful than the parametric approaches comparing the LR values (e.g., t-test), when the assumption of normality satisfies. However, for complex disease models involving interactions, such assumption may not hold. When the assumption of normality is violated, non-parametric approaches, such as TAMW, could be powerful than the parametric approaches[Blair and Higgins 1981; Blair, et al. 1980].

Although TAMW has several advantages over existing approaches, it also has certain limitations. TAMW, likes many ensemble learning approaches, does not explicitly model the interaction by estimating interaction effect sizes, which makes the interaction less interpretable. Moreover, the tradeoff to using the computationally efficient forward selection algorithm in TAMW is that at least one of the selected loci must have a reasonably strong marginal effect. When interaction exists among loci in the complete absence of main effects, the approach could have lower power as compared to an exhaustive search approach, MDR (Figure 1). The TAMW described in this paper does not consider covariate adjustment. The easiest way to handle covariates in the TAMW is to build two models, one with covariates and the other without covariates. By comparing these two models, we are able to test the joint association, adjusting for covariate effects. Alternatively, we could build propensity scores[Jiang and Zhang 2011] or stratification scores[Allen and Satten 2011] on the covariates, and then use the scores as weights in the TAMW approach for covariate adjustment.

## Acknowledgements

We appreciate critical input from two anonymous reviewers. This study makes use of data generated by the Wellcome Trust Case Control Consortium. Funding for the original WTCCC project was provided by the Wellcome Trust under award 076113. A full list of the investigators who contributed to the generation of the data is available from http://www.wtccc.org.uk/info/participants.shtml. This work was supported by the National Institute of Dental and Craniofacial Research under Award Number R03DE022379. The content is solely the responsibility of the authors and does not necessarily represent the official views of the National Institutes of Health.

# Figure

**Figure 1: Power comparison of LRMW, TAMW and MDR under five interaction models**

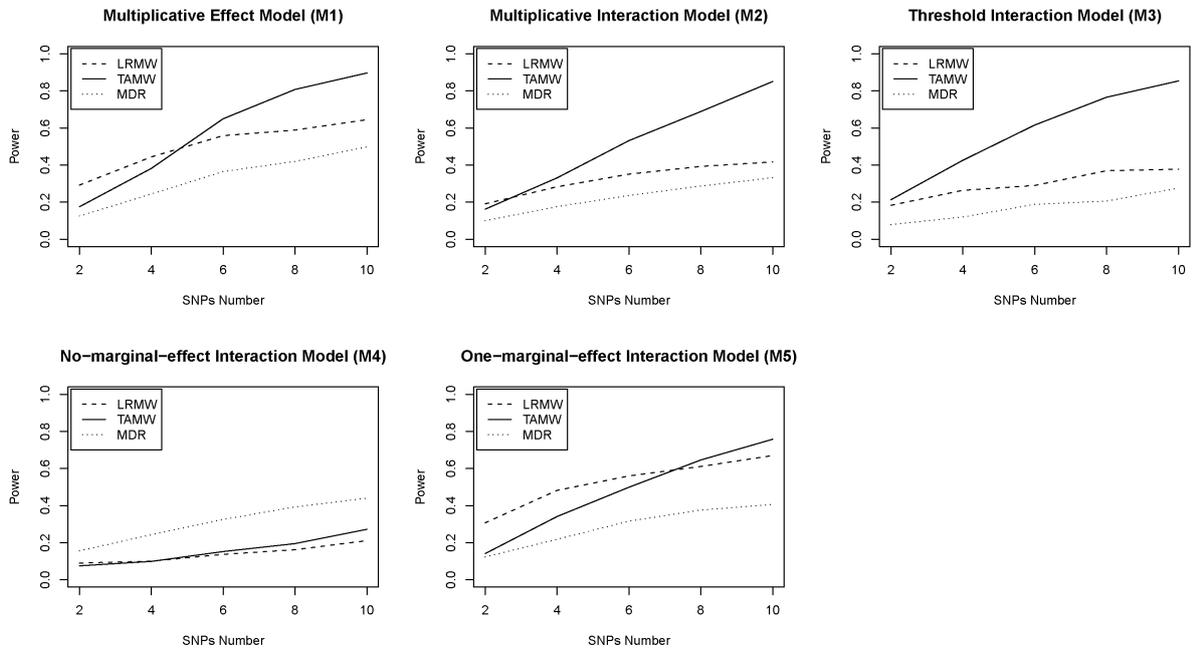

# Tables

**Table I: Power comparisons for TAMW, MDR and LRMW**

| Method | Power | | | | | |
|---|---|---|---|---|---|---|
| | 0 SNPs | 2 SNPs | 5 SNPs | 10 SNPs | 15 SNPs | 20 SNPs |
| **TAMW** | 0.053 | 0.691 | 0.761 | 0.744 | 0.83 | 0.931 |
| **MDR** | 0.024 | 0.705 | 0.570 | 0.434 | 0.462 | 0.599 |
| **LRMW** | 0.054 | 0.812 | 0.771 | 0.699 | 0.686 | 0.704 |

**Table II: Summary of the candidate gene analysis**

| Chr. | Gene | SNPs | TAMW | MDR | LRMW |
|---|---|---|---|---|---|
| **1p31** | *IL23R* | rs11465804 | 1[a] | +[b] | 3[c] |
| **2q37** | *ATG16L1* | rs3828309 | 2 | + | 2 |
| **5q33** | | rs11747270 | 3 | | |
| **18p11** | | rs2542151 | 4 | | |
| **13q14** | *LACC1* | rs3764147 | 5 | | |
| **5p12** | | rs4613763 | 6 | + | 1 |
| **9q32** | *TNFSF15* | rs4263839 | 7 | | |
| **1q25** | | rs9286879 | 8 | | |
| **9q24** | | rs10758669 | 9 | | |
| **10q24** | *NKX2-3* | rs11190140 | 10 | | |
| | | others | 11~29 | | |
| **p-value** | | | 1.84e-39 | 5.81e-05 | 6.25e-14 |

[a] Rank of IM value from TAMW.

[b] SNPs selected by MDR.

[c] Number represents the selection sequence from LRMW.

**Table III: Summary of GWAS analysis using TAMW**

| Chr. | Gene | SNPs | IM | Rank | Single-locus P-value | Single-locus rank |
|---|---|---|---|---|---|---|
| 1p31 | IL23R | rs2201841 | 3.38 | 12 | 6.51e-12 | 14 |
| | | rs11805303 | 2.90 | 26 | 8.11e-13 | 7 |
| | | rs10489629 | 2.45 | 54 | 6.05e-12 | 13 |
| | | rs6664119 | 2.19 | 84 | 7.54e-06 | 105 |
| | IL12RB2 | rs3790567 | 2.75 | 35 | 1.84e-05 | 133 |
| 1q41 | GALNT2 | rs12751815 | 2.31 | 69 | 4.59e-04 | 651 |
| 2q37 | ATG16L1 | rs3792106 | 4.20 | 1 | 1.69e-11 | 19 |
| | | rs10210302 | 4.07 | 2 | 9.10e-14 | 2 |
| | | rs6752107 | 4.03 | 3 | 1.56e-13 | 4 |
| | | rs6431654 | 3.88 | 4 | 1.11e-13 | 3 |
| | | rs3828309 | 3.80 | 5 | 4.44e-13 | 6 |
| 3q27 | MCF2L2 | rs2314349 | 2.61 | 44 | 4.35e-04 | 624 |
| 5q31 | C5orf56 | rs4540166 | 2.94 | 25 | 9.40e-06 | 110 |
| | SLC22A5 | rs274547 | 2.16 | 88 | 7.65e-07 | 73 |
| | | rs274552 | 2.73 | 36 | 9.49e-07 | 79 |
| 9q32 | TNFSF15 | rs6478108 | 2.42 | 59 | 9.20e-05 | 261 |
| 13q14 | LACC1 | rs3764147 | 2.23 | 77 | 4.10e-05 | 183 |
| 16q12 | NOD2 | rs17221417 | 2.91 | 27 | 1.14e-11 | 17 |
| | | rs3135499 | 2.82 | 30 | 1.21e-08 | 28 |
| | | rs2076756 | 2.46 | 52 | 8.35e-15 | 1 |
| | | rs2066843 | 2.27 | 73 | 1.48e-12 | 8 |
| | | rs1861759 | 2.47 | 51 | 8.01e-07 | 74 |
| | | rs748855 | 2.69 | 40 | 2.78e-07 | 63 |
| | CYLD | rs7342715 | 2.77 | 32 | 6.59e-09 | 26 |
| | | rs11076540 | 2.76 | 34 | 2.87e-08 | 39 |
| | | rs3135503 | 2.40 | 61 | 4.56e-08 | 45 |
| 18p11 | PTPN2 | rs16939895 | 3.07 | 17 | 5.88e-07 | 71 |
| | | rs7234029 | 2.90 | 28 | 4.06e-07 | 68 |
| | | rs3135503 | 2.40 | 61 | 4.56e-08 | 45 |
| 20q13 | ZGPAT | rs6011040 | 2.50 | 47 | 9.12e-05 | 260 |
| | | rs2738758 | 2.10 | 94 | 1.95e-05 | 137 |